\documentclass[11pt,a4paper]{article}
\pdfoutput=1
\usepackage{amsmath,amssymb,amsfonts,a4wide,graphicx,bm,times}
\usepackage{mcite}
\makeatletter \let\old@startsection=\@startsection
\renewcommand{\@startsection}[6]
{\old@startsection{#1}{#2}{#3}{#4}{#5}{#6\mathversion{bold}}}
\makeatother

\makeatletter

\@addtoreset{equation}{section}
\makeatother
\let\refOld\ref
\renewcommand{\ref}[1]{(\refOld{#1})}

\newcommand{\tr}{{\rm tr~}}

\newcommand{\superp}[2]{\genfrac{}{}{0pt}{}{#1}{#2}}

 \def\d{\delta}

 \def\p{\partial}
 
 \def\a{\alpha}
 \def\b{\beta}
 \def\g{\gamma}
 \def\d{\delta}
 \def\e{\epsilon}

 \def\l{\lambda}

 \def\t{\tau}

 \def\G{\Gamma}

 \def\O{\Omega}

\def\equskip{\!\!\!\!\!\!\!\!} 
\def\la{\left\langle}
\def\ra{\right\rangle}

\def\hf{\dfrac{1}{2}}

\def\implies{\quad\Rightarrow\quad}

\def\mut{\tilde{\mu}}

\def\CF{\mathcal{F}}

\def\CA{{\mathcal{A}}}
\def\CZ{{\mathcal{Z}}}
\def\CB{{\mathcal{B}}}

\def\CZb{\mathcal{Z}_{\b\text{-ens}}}
\def\bens{$\b$-ensemble}
\def\tH{\tilde{H}}

\begin{document}
\begin{titlepage}
\renewcommand{\thefootnote}{\fnsymbol{footnote}}

\vspace*{1cm}
    \begin{Large}
       \begin{center}
         {Large $N$ limit of $\b$-ensembles and deformed Seiberg-Witten relations}
       \end{center}
    \end{Large}
\vspace{0.7cm}

\begin{center}
Jean-Emile B{\sc ourgine}\footnote
            {
e-mail address : 
jebourgine@sogang.ac.kr}\\
      
\vspace{0.7cm}                    
{\it Center for Quantum Spacetime (CQUeST)
}\\
{\it Sogang University, Seoul 121-742, Korea}
\end{center}

\vspace{0.7cm}

\begin{abstract}
\noindent
We study the \bens\ that represents conformal blocks of Liouville theory on the sphere. This quantity is related through AGT conjecture to the Nekrasov instanton partition function of 4d $\mathcal{N}=2$ $SU(2)$ gauge theory with four flavors. We focus on the large $N$ limit, equivalent to the Nekrasov-Shatashvili limit where one of the $\O$-background deformation parameters is vanishing. A quantized Seiberg-Witten differential form is defined perturbatively in $\hbar$ as the singular part of the \bens\ resolvent. Using the Dyson collective field action, we show that the free energy obeys the Seiberg-Witten relations. As suggested by Mironov and Morozov, the quantized differential form can be obtained from the classical one by the action of a differential operator in the hypermultiplet masses and the Coulomb branch modulus.

\end{abstract}
\vfill

\end{titlepage}
\vfil\eject

\setcounter{footnote}{0}

\section{Introduction}

Nekrasov and Shatashvili proposed in \cite{Nekrasov2009} to quantize the correspondence between the Seiberg-Witten low energy effective theory of 4d $\mathcal{N}=2$ supersymmetric gauge theories \cite{Seiberg1994,*Seiberg1994a}, and integrable models characterized by a complex curve, the Hitchin systems \cite{Gorsky1995}. The Yang-Yang function of the quantum integrable model \cite{Yang1968} is associated to a $\b$-deformed prepotential given by the Nekrasov instanton partition function of the gauge theory on the $\O$-background with one vanishing deformation parameter $\e_2=0$. The second deformation parameter $\e_1$ is then identified with the Planck constant. When this second parameter is also vanishing, the instanton partition function reduces to the Seiberg-Witten prepotential of the gauge theory on $\mathbb{R}^4$ \cite{Nekrasov2003,Nekrasov2003a}. Based on this correspondence, it has been proposed in \cite{Mironov2010,*Mironov2010a,*Popolitov2010} to compute the quantized prepotential perturbatively in $\hbar$ from Bohr-Sommerfeld integrals appearing in the semi-classical treatment of the quantum mechanical system. This prepotential was then successfully compared to the instanton partition function at first orders.

On the other hand, the celebrated Alday-Gaiotto-Tashikawa (AGT) conjecture \cite{Alday2009} relate the instanton partition function of 4d $\mathcal{N}=2$ SUSY gauge theories with a product of $SU(2)$ gauge groups to a correlator of Liouville theory defined on a Riemann surface.\footnote{The correspondence was later generalized to $SU(n)$ gauge groups and Toda field theory \cite{Wyllard2009,*Bonelli2009,Mironov2010b}.} This conjecture has been checked in a number of specific limiting cases \cite{Mironov2009a,*Giribet2009,*Marshakov2009a,*Nanopoulos2009,*Alba2009,*Marshakov2009b,*Belavin2011,*Kanno2011,Mironov2009}, and even proven for a single $SU(2)$ gauge group and $N_f\leq2$ flavors \cite{Fateev2009,*Hadasz2010}. We focus here on the $\mathcal{N}=2$ gauge theory with a $SU(2)$ gauge group and four scalar matter hypermultiplets in the fundamental representation. This theory is known to be scale invariant even at the non-perturbative level, and is sometimes referred to as superconformal QCD. The instanton contribution to the partition function is related to the Liouville conformal block, and the full partition function, integrated over the moduli space, to the Liouville four points function. Inspired by the CFT methods developed for matrix models in \cite{Marshakov1991,*Kharchev1992} (see also \cite{Kostov1999,*Kostov2009,*Kostov2010}), Dijkgraaf and Vafa \cite{Dijkgraaf2009} suggested to realize the Liouville conformal block as a $\b$-deformed matrix model, the $\b$-ensemble. This proposal was soon generalized to higher genus and various gauge groups \cite{Itoyama2009,Bonelli2011,*Maruyoshi2011,*Schiappa2009} and thoroughly investigated \cite{Mironov2009b,Mironov2010c,Mironov2010e,*Itoyama2010a,*Itoyama2011}. 

The proposal exposed in \cite{Dijkgraaf2009} reformulates the AGT conjecture as a relation between the \bens\ integrals and the Nekrasov partition function. This identity has been proved in \cite{Mironov2010d} for the case $\b=1$ (Hermitian matrix model) which corresponds to the relation $\e_1=-\e_2$ between the gauge theory deformation parameters.\footnote{This demonstration has been generalized to an $SU(n)$ gauge group in \cite{Zhang2011}.} In the large $N$ limit, where $N$ is the size of the matrix, the \bens\ reproduces the Seiberg-Witten theory \cite{Eguchi2010,Eguchi2010a}, and the Seiberg-Witten curve in the Gaiotto form \cite{Gaiotto2009a} is recovered as the spectral curve of the matrix model.

The instanton partition function with $\e_2=0$ is believed to satisfy $\hbar$-deformed Seiberg-Witten relations \cite{Nekrasov2009,Mironov2010}. Following the Dijkgraaf-Vafa formulation of the AGT conjecture, the free energy of the \bens\ should also satisfies the same relations in the large $N$, or planar, limit. Until now, this claim has only been verified for $\b=1$ \cite{Chekhov2002}. One of the main purpose of this paper is to extend this demonstration using the Dyson collective field representation of the planar free energy \cite{Dyson1962}. When $\e$-deformation parameters are turned on, the differential form of Seiberg-Witten theory has to be modified. It is largely believed that the quantized differential form is given by the resolvent of the \bens\ as $H(z)dz$ \cite{Mironov2011a}. Our study of the collective field action shows that this statement is slightly incorrect, and one has to withdraw a regular function from $H(z)$ to recover the differential form $\p \eta(z)dz$. This regular term is usually discarded from the cycle integrals of the Seiberg-Witten relations as a total derivative. However, it contains poles at the branch points which gives a non-zero contribution, as already noticed in \cite{Nishinaka2011}. Such poles do not pertain to the singularities of the logarithmic potential but are a general feature of the WKB expansion. The discrepancy between the resolvent and the differential form is vanishing at $\hbar=0$, it is related to the cut-off term of the Dyson action. Moreover, the resolvent $H(z)$ is the classical $U(1)$ current in the free field representation of the \bens, and the relation \ref{decompo_H} between $H(z)$ and $\eta(z)$ is the transformation law of this current under the conformal mapping $\eta\to z(\eta)$. Thus, $\eta(z)$ can be interpreted as the change of coordinates which trivializes this classical current.

Superconformal QCD with $SU(2)$ gauge group is characterized by a four punctured Riemann sphere in the M-branes construction \cite{Gaiotto2009a}. In this simple case, the associated quantum Hitchin system takes the form of a Schr\"odinger equation \cite{Maruyoshi2010,Bonelli2011a}. The quantized differential form can be defined from the solution of this Schr\"odinger equation, and computed explicitly in the WKB expansion. In \cite{Mironov2010,*Mironov2010a} was suggested the existence of a differential operator in the parameters of the Seiberg-Witten curve (mass of the hypermultiplets, Coulomb branch modulus, gauge coupling) generating the higher order terms in the $\hbar$-expansion of the deformed differential $\p \eta(z)dz$ from the $\hbar$-classical differential. The existence of such a structure has been observed at first order in $\hbar$ in \cite{Maruyoshi2010}, and will be demonstrated to hold at all orders.

The restriction to the gauge theory with four massive hypermultiplets here is mostly for simplicity, and we believe our results extend to theories with a smaller number of flavors. These theories may indeed be obtained by sending some of the hypermultiplet masses to infinity, a procedure which has an equivalent in Liouville theory and \bens\ \cite{Gaiotto2009b,*Marshakov2009,*Itoyama2010}. We should more particularly emphasize that the demonstration of the planar free energy subjection to the Seiberg Witten relations does not depend on the explicit form of the potential. Extending the proof to higher orders in the large $N$, or topological, expansion is a challenging issue that may be addressed using the topological recursion tools. This matrix model technique recently extended to the \bens s \cite{Chekhov2006,Chekhov2010a,Eynard2008a,*Chekhov2009,*Chekhov2010} allows to compute the full free energy of the model which is identified with the $(\e_1,\e_2)$-deformed prepotential. Whether this quantity satisfies the Seiberg-Witten relations is still an open question. 

It would be interesting to consider further generalizations to higher gauge groups and quiver \bens s. Another promising direction is the relation with the direct formulation of the Nekrasov partition function as a \bens\ \cite{Klemm2008,*Sulkowski2009}. Finally, one of the most intriguing open questions remains to find a connection with the deformed Seiberg-Witten curve introduced in \cite{Poghossian2010,*Fucito2011,*Huang2012}, and derived with the method of Young tableaux profile density developed in \cite{Nekrasov2003a}. To do so, we need a definition of the quantized curve which is independent of the quantization scheme. This may be given by the underlying integrable models \cite{Mironov2010,Teschner2012,Zenkevich2011}, or could also arise from a geometrical point of view \cite{Reffert2011,*Hellerman2011,*Hellerman2012}.

\section{Planar limit of the $\b$-ensemble}
\subsection{Definition of the model}
To formulate the AGT conjecture on the \bens\ partition function, we start from the Liouville correlator of four vertex operators inserted at the points $q_f=0,1,q,\infty$ and with respective charge $\a_f=\a_0,\a_1,\a_2,\a_\infty$. We further assume that the charges satisfy the neutrality condition
\begin{equation}\label{neutral}
\sum_{f=0,1,2,\infty}{\a_f}+Nb=Q_L,
\end{equation}
where $N$ is a positive integer and $Q_L=b+1/b$ is the Liouville background charge. Following \cite{Alday2009}, we shift two of the charges in order to absorb the background charge $Q_L$, $\a_0\to\a_0+Q_L/2$ and $\a_\infty\to \a_\infty+Q_L/2$. This manipulation corresponds to concentrate the curvature of the sphere equally at zero and infinity. Since we are interested in a semi-classical expansion of heavy fields \cite{Zamolodchikov1996}, we introduce a scale $g$ and the rescaled quantities $m_f=-2ig\a_f$, $\hbar=-ig Q_L$.\footnote{Our case differ from \cite{Mironov2009} where the AGT conjecture has been proved in the semi-classical limit $c\to\infty$ (or $\e_2=0$) but only for light fields, i.e. $m_f=O(\e_2)$.} This scale $g$ plays the role of the Planck constant for Liouville theory, and we will later consider the expansion $g\to0$, $N\to\infty$ at fixed $gN,\hbar$ and $m_f$. In these new variables, the condition \ref{neutral} is rewritten
\begin{equation}\label{neutral_mf}
\sum_{f=0,1,2,\infty}{m_f}+2\sqrt{\b}gN=0,\qquad (\text{with }b=i\sqrt{\b}).
\end{equation}

After integration over the zero mode, the Liouville four points correlation function exhibits a pole when the relation \ref{neutral} is satisfied. The residue can be expressed as a free field correlator with $N$ screening charges inserted \cite{Goulian1990}. The conformal blocks entering in the decomposition of such correlators have a well known integral representation which was studied by Dotsenko and Fateev \cite{Dotsenko1984,*Dotsenko1984a}. According to the proposal of Dijkgraaf and Vafa \cite{Dijkgraaf2009}, further studied in \cite{Mironov2009b,Mironov2010c,Mironov2010e}, these integrals can be written as a \bens\ partition function. The crucial issue is the specification of the integration contours for the screening operators, it has been resolved in the perspective of the AGT conjecture. This choice is believed to be equivalent to the quasiclassical approach to \bens s we use in this paper, as shown at first orders in \cite{Morozov2010}.

Using the \bens\ representation for the Liouville conformal block, the AGT conjecture reads
\begin{equation}\label{AGT}
\CZb(a,m_i,g,\b)=q^{-((m_0-m_2)^2+4\hbar(m_0-\hbar))/4g^2}(1-q)^{2(m_1-2\hbar)(m_2-2\hbar)/4g^2}\mathcal{Z}_\text{Nekrasov}(a,\mu_i,\e_\a),
\end{equation}
up to a $q$-independent constant factor that can be found in \cite{Mironov2010c}.\footnote{We consider here the \bens\ used in \cite{Eguchi2010,Nishinaka2011}. To obtain the correct powers of $q$ and $q-1$ from \cite{Alday2009}, one should take into account the flip $a\to-a$, $m_0\to-m_0$, $m_\infty\to-m_\infty$, $m_1\to2\hbar-m_1$ and $m_2\to2\hbar-m_2$ that leaves the $SU(2)$ instanton partition function invariant. This flip corresponds to the action of the Weyl group of $SU(2)$ gauge and flavor symmetry, and to the reflexion $\a\to Q_L-\a$ of Liouville vertex operators.} On the RHS, $\mathcal{Z}_\text{Nekrasov}$ refers to the full gauge theory partition function, including classical, one-loop and instantons contributions. It depends on the UV gauge coupling $\t$ through the parameter $q=e^{2i\pi\t}$ identified with the cross-ratio of the conformal block. On the LHS, the $\b$-ensemble partition function $\CZb$ is a multiple integral over $N$ variables, the eigenvalues $\l_I$, with a potential $V(\l)$ and a Van der Monde determinant at the power $2\b$,
\begin{equation}\label{def_beta_ens}
\CZb(a,m_i,g,\b)=\int{\prod_{I=1}^N{e^{\frac{\sqrt{\b}}{g}V(\l_I)}d\l_I }\prod_{\superp{I,J=1}{I\neq J}}^{N}{\left(\l_I-\l_J\right)^{\b}}}.
\end{equation}
When $\b=1$, we recognize the eigenvalue integrals appearing in the Hermitian matrix model after integrating out the angular degrees of freedom, and the model \ref{def_beta_ens} is sometimes also referred as a ``$\b$-deformed'' matrix model. The ``Penner type'' potential $V$ contains logarithmic terms taken at the finite insertion points of the vertex operators,
\begin{equation}\label{Penner}
V(z)=(m_0+\hbar)\log z+m_1\log(z-1)+m_2\log(z-q).
\end{equation}
This potential is independent of the parameter $m_\infty$ which appears in \ref{def_beta_ens} only through the matrix size $N$ following the relation \ref{neutral_mf}. The masses $\mu_i$ of the four matter hypermultiplets are related to the charge of the vertex operators through
\begin{align}
\begin{split}
&\mu_1=m_1+m_\infty,\qquad\qquad \mu_3=m_2+m_0,\\
&\mu_2=m_1-m_\infty,\qquad\qquad \mu_4=m_2-m_0,
\end{split}
\end{align}
and with a slight abuse of terminology we call the parameters $m_f$ ``masses'' associated to the insertion points $q_f=0,1,q,\infty$. The equivariant deformation parameters $\e_1$ and $\e_2$ are connected to the $\b$-ensemble quantities as
\begin{equation}
g^2=-\dfrac14\e_1\e_2,\quad \b=-\dfrac{\e_1}{\e_2}=-b^2,\quad \hbar=\dfrac{g}{\sqrt{\b}}(\b-1)=\hf(\e_1+\e_2).
\end{equation}
Finally, the Coulomb branch vev of the gauge theory $a$ is related to the momentum $\a$ of the conformal block in the intermediate channel as $a-\hbar=2ig\a$. In the quasiclassical approach, the eigenvalues of the \bens\ condense on a continuum of saddle points neighboring the minima of the potential, leading to an eigenvalue density supported on several intervals. The potential \ref{Penner} defines a model with two such intervals. These are the two branch cuts of the resolvent, the Cauchy transform of the eigenvalue density. The filling fraction $a$ describes the ratio of the eigenvalues that occupy one of these branch cuts. More precisely, this filling fraction condition will be later interpreted as one of the deformed Seiberg-Witten relations.\\

The deformation parameters have dimension of a mass, and the Coulomb branch vev $a$ inherits the dimension of the scalar field in the gauge multiplet. The conformal invariance of the gauge theory implies the invariance under the rescaling $\mu_i\to\g\mu_i$, $a\to\g a$ and $\e_{1,2}\to\g\e_{1,2}$. \footnote{The gauge theory side scale invariance should not be confused with the Liouville theory scale invariance which acts on the insertion position of the vertex operators and has already been used to fix these positions at $0,1,q,\infty$.} This transformation allows to fix one of the $\e$-parameters to one, so that the theory depends only on the ratio $\b$. Similarly, the Liouville theory only depends on the parameter $b=i\sqrt{\b}$ and the scale $g$ has been introduced by hand. After a proper rescaling, \bens\ correlators only depend on $\b$ through the parameter $\hbar$.\footnote{This property can be used to fix $\b$ upon a change in the matrix size which controls the value of $g$. At $\b=1/2$ (or $\b=2$) the partition function is identified with the eigenvalue integrals of an orthogonal (or simplectic) matrix model. The limit $\b\to1$ with $\hbar$ fixed corresponds to send $g$ to infinity, or the matrix size toward zero.} Again, the parameters $(g,\hbar)$ are not independent but obey the invariance under rescaling $g\to\g g$, $\hbar\to\g\hbar$ together with $m_f\to\g m_f$ and $a\to\g a$.

In the large $N$ limit of the \bens, we send $g\to0$ so that the fixed product $gN$ obeys the relation \ref{neutral_mf}, and the large $N$ topological expansion is an expansion at small $g$. There are two possibilities to define this expansion, considering either $\b$ or $\hbar$ fixed. Due to the scale invariance mentioned above, it seems more natural in the context of AGT to fix $\b$ so that $\hbar=O(g)$. However, it reveals more convenient here to keep the two variables ($g,\hbar$) independent and take the double expansion of the $\b$-ensemble quantities. In this paper, we should focus on the zeroth order in $g$, but all orders in $\hbar$, which we refer as the ``planar limit''. This limit is equivalent to the Nekrasov-Shatashvili limit \cite{Nekrasov2009} where one of the deformation parameters is set to zero, while the second one is identified with the Planck constant $2\hbar$. The remaining $\hbar$ expansion of planar quantities is interpreted as a WKB expansion. The planar limit also coincides with the semi-classical limit of Liouville theory in which $c\to\infty$.

Another interesting limit is $\hbar\to0$ with fixed $g$, i.e. $\b=1$, where the $\b$-ensemble is identified with an Hermitian matrix model. It corresponds to set $\e_1=-\e_2$ for the $\O$-background, and gives the $c\to 1$ limit of Liouville theory ($Q_L\to0$). In this limit, the AGT relation \ref{AGT} between the instanton partition function and the \bens\ has been demonstrated \cite{Mironov2010d}. Combining both limiting cases, we reach the planar limit of an Hermitian matrix model at $g=\hbar=0$. It corresponds to consider the gauge theory on the flat $\mathbb{R}^4$ background. On the Liouville side of the correspondence, we get the semi-classical limit of a $c=1$ free boson.

We define the planar free energy and the deformed Seiberg-Witten prepotential as
\begin{equation}
\CF_\b=\lim_{g\to0}{g^2\log \CZ_\b},\qquad \CF_\text{SW}=-\lim_{\e_2\to0}{\e_1\e_2\log\CZ_\text{Nekrasov}},
\end{equation}
and in the planar limit the AGT relation \ref{AGT} reduces to
\begin{equation}\label{AGT_planar}
\CF_\text{SW}=4\CF_\b+\left((m_0-m_2)^2+4\hbar (m_0-\hbar)\right)\log q-2(m_1-2\hbar)(m_2-2\hbar)\log(q-1)+C
\end{equation}
with $\e_1=2\hbar$ and a $q$-independent constant $C$. This constant is also believed to be independent of $a$. The relation \ref{AGT_planar} was shown to hold up to the order $O(\hbar,q^4)$ at the special point $m_0=m_\infty=0$, $m_1=m_2$ in \cite{Nishinaka2011} and this result has been extended very recently to the order $O(\hbar^2)$ \cite{Nishinaka2012}.

\subsection{Loop equation and Schr\"odinger equation}
Over the past fifty years, matrix models have been thoroughly investigated and various methods have been developed to solve them. Unfortunately, only a part of these techniques have been successfully extended to \bens s \cite{Morozov2012a,*Mironov2012,*Morozov2012}. Nonetheless, a $\b$-deformed version of the topological recursion has been developed in \cite{Chekhov2006,Chekhov2010a}. These algebraic geometry methods provide a formal recursive solution to the loop equations that still needs to be analyzed in the framework of the AGT conjecture. In this context, we believe that the methods and ansatz used here in the planar limit have a straightforward generalization to higher orders.

Loop equations are derived by exploiting the invariance of the measure under a change of variables. The simplest equation is obtained from the shift
\begin{equation}
\l_I\to\l_I+\dfrac{\e}{z-\l_I},
\end{equation}
at the order $O(\e)$ and has been derived in \cite{Eguchi2010a} for the Penner type model \ref{Penner}. It writes at the first order in $g$,
\begin{equation}\label{le1}
W(z)^2+\hbar W'(z)+V'(z)W(z)-U(z)=0,
\end{equation}
where we defined the planar resolvent $W$ and the auxiliary correlator $U$ as
\begin{equation}\label{def_resol}
W(z)=\lim_{g\to0}g\sqrt{\b}\sum_{I=1}^N{\la\dfrac1{z-\l_I}\ra},\quad U(z)=\lim_{g\to0}g\sqrt{\b}\sum_{I=1}^N\la\dfrac{V'(z)-V'(\l_I)}{z-\l_I}\ra.
\end{equation}
The quantity $U(z)$ is a sum of simple poles at $z=q_i$ where $q_0=0,\ q_1=1,\ q_2=q$, with residues proportional to $W(q_i)$. These three quantities are not independent but are related through two loop equations derived from the change of variable $\l_I\to\l_I+\e$ and $\l_I\to\l_I+\e\l_I$ respectively,
\begin{equation}\label{sym_rel_I}
\sum_{i=0}^2{(m_i+\d_{i0}\hbar)W(q_i)}=0,\quad 4\sum_{i=0}^2{q_i(m_i+\d_{i0}\hbar)W(q_i)}=m_\infty^2-\left(\sum_{i=0}^2{m_i}\right)^2.
\end{equation}
These relations can also be obtained from the asymptotic of \ref{le1} at large $z$, noting that $W(z)\sim \sqrt{\b}gN/z$. They allow to express $U(z)$ in terms of the mass parameters, the cross-ratio $q$, and the quantity $u=qm_2W(q)$. An interesting feature of the Penner-type potential \ref{Penner} is that the resolvent evaluated at the insertion point $z=q$ is simply the derivative of the free energy with respect to this parameter. Combining this property with the planar AGT relation \ref{AGT_planar}, we may write
\begin{equation}\label{deriv_q}
q\dfrac{\p\CF_\text{SW}}{\p q}=4u+(m_0-m_2)^2+4\hbar (m_0-\hbar)-2(m_1-2\hbar)(m_2-2\hbar)\dfrac{q}{q-1},\quad u=q\dfrac{\p\CF_\b}{\p q}.
\end{equation}
The comparison between the matrix model at $\b=1$ and the Seiberg-Witten theory made in \cite{Eguchi2010a} relates the quantity $u$, up to a mass-dependent translation, to the Coulomb branch parameter $\la\tr\Phi^2\ra$ where $\Phi$ is the scalar field in the $\mathcal{N}=2$ gauge multiplet.\footnote{The precise identification with the results of \cite{Eguchi2010a} involves the change of notations from here to there: $m_0\to m_1$, $m_1\to m_2$, $m_2\to m_3$, $m_\infty\to m_0$ and $u$ is related to their quantity $U$ as
\begin{equation}
(q+1)U=-4u(q-1)+qm_\infty^2-(2q-1)m_2^2-qm_1^2+m_0^2-4m_1m_2-2(q-1)m_0m_2
\end{equation}
in our notations.} In the Seiberg-Witten theory, it is usual to consider $u$ as an independent parameter. On the other hand, $u$ being proportional to the planar resolvent taken at $z=q$, it has a natural expansion in $\hbar$. The relation between these two points of view, $u$ fixed ($a$ function of $u$ and $\hbar$) and $u$ function of $a$ (fixed) and $\hbar$, will be investigated in the last section, and for now we adopt the Seiberg-Witten perspective. In addition, to simplify the quantities, we shift the variable $u$ into
\begin{equation}\label{def_E}
E=4u+2m_2(m_0+2\hbar)+2m_1m_2\dfrac{q}{q-1}-m_0^2-m_2^2.
\end{equation}
With a slight abuse of terminology, we will refer to $E$ as the Coulomb modulus parameter of the gauge theory.

Introducing
\begin{equation}\label{def_HT}
H(z)=W(z)+\hf V'(z),\quad T(z)=U(z)+\dfrac14V'(z)^2+\dfrac{\hbar}{2}V''(z),
\end{equation}
the planar loop equation \ref{le1} takes the form of a Riccati equation,
\begin{equation}\label{equa_diff}
H(z)^2+\hbar\p H(z)=T(z).
\end{equation}
The meromorphic function $T(z)$ has double poles at $z=q_i$ and behaves as $O(1/z^2)$ at infinity. It has a linear dependence in $E$,
\begin{equation}
T(z)=\dfrac{p_4(z)}{\prod_{i=0}^2{(z-q_i)^2}}+\dfrac{(q-1)E}{4\prod_{i=0}^2{(z-q_i)}},
\end{equation}
where $p_4(z)$ is a polynomial of degree four in $z$ which is quadratic in $\hbar$, it can be expanded as
\begin{align}
\begin{split}\label{expr_poly}
p_4(z)&=p_4^{(0)}(z)+\hbar p_4^{(1)}(z)+\hbar^2 p_4^{(2)}(z),\\
4p_4^{(0)}(z)&=m_\infty^2z^4+\left(q(-2m_\infty^2+3m_2^2+m_0^2)-m_\infty^2-2m_2^2+m_1^2-2m_0^2\right)z^3\\
&+\left(q^2(m_\infty^2-2m_2^2-m_0^2)+q(2m_\infty^2-2m_2^2-2m_1^2+2m_0^2)+2m_2^2+2m_0^2\right)z^2\\
&+\left(q^2(-m_\infty^2+2m_2^2+m_1^2)-q(m_2^2+3m_0^2)\right)z+m_0^2q^2\\
p_4^{(1)}(z)&=-\left(\dfrac{\p}{\p m_1}+\dfrac{\p}{\p m_2}\right)p_4^{(0)}(z),\quad 4p_4^{(2)}(z)=-(z-1)^2(z-q)^2.
\end{split}
\end{align}

It is well known that the Riccati equation \ref{equa_diff} can be cast as a Schr\"odinger equation for the wave function $\psi$ defined as $H=\hbar\p\log\psi$,
\begin{equation}\label{Schrodinger}
\left(\hbar^2\p^2-\dfrac{p_4(z)}{\prod_{i=0}^2{(z-q_i)^2}}\right)\psi(z)=\dfrac{(q-1)E}{4\prod_{i=0}^2{(z-q_i)}}\psi(z).
\end{equation}
This one dimensional quantum mechanical problem is the integrable system associated to the Riemann sphere with four punctures at $z=q_f$ \cite{Maruyoshi2010,Tai2010}. Up to the pole factor, the parameter $E$ is seen as the energy of the system, whereas the ratio in the LHS is identified with the potential. In Liouville theory, the wave function $\psi(z)$ is constructed as the insertion of the degenerate operator $\phi_{2,1}(z)$ within the four point correlator. The Schr\"odinger equation \ref{Schrodinger} is a consequence of the null state condition of level two obeyed by this operator. In the \bens, the wave function is realized as the planar limit of the following correlator,
\begin{equation}
\psi(z)=e^{\frac1{2\hbar}V(z)}\lim_{g\to0}\la\prod_{I=1}^N(z-\l_I)^{\frac{\sqrt{\b}g}{\hbar}}\ra.
\end{equation}
This quantity is associated through the AGT correspondence to the vev of a surface operator in the gauge theory \cite{Alday2009a}.

The appearance of the Schr\"odinger equation is at the origin of the ``quantum spectral curve'' notion defined for the $\b$-ensemble in \cite{Eynard2008a,*Chekhov2009,*Chekhov2010}. In this context, two non-commuting variable $x$ and $y$ satisfying $[y,x]=\hbar$ are introduced. The spectral curve is the rational fraction $\mathcal{E}(x,y)=y^2-T(x)$ and act on the wave functions $\psi(x)$ as
\begin{equation}
\mathcal{E}(x,y)\psi(x)=0,\quad y=\hbar\dfrac{\p}{\p x}.
\end{equation}
At finite $\hbar$, the branch cuts of the resolvent disappear, they are replaced by half-line of accumulation of zeros for $\psi$. Here, we take a different approach and keep $\hbar$ infinitesimal, using the WKB expansion to solve the Schr\"odinger equation. At each order in $\hbar$ the resolvent keeps the same branch cuts, which is suited for a comparison with a $\b$-deformed Seiberg-Witten theory. 

\subsection{Classical (or Hermitian) limit}
\begin{figure}
\begin{center}
\includegraphics[width=10cm]{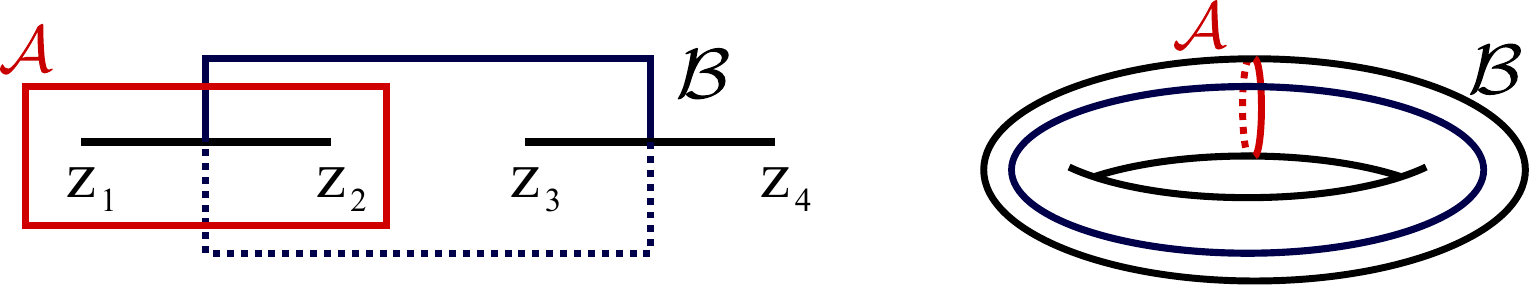}
\caption{Definition of the $\CA$ and $\CB$ cycles on the spectral curve.}\label{fig1}
\end{center}
\end{figure}

The WKB solution consists in expanding the functions in $\hbar$ to solve the problem recursively at each order,
\begin{equation}\label{WKB}
H(z)=\sum_{n=0}^\infty{\hbar^nH_n(z)},\quad T(z)=\sum_{n=0}^2{\hbar^nT_n(z)}.
\end{equation}
At the zeroth order, we recover the planar limit of an Hermitian matrix model. The equation for the critical resolvent $H_0(z)$ defines the spectral curve $(H_0,z)\in\mathbb{C}\times\mathbb{C}$ of the model,
\begin{equation}\label{spec_curve}
H_0(z)^2=\dfrac{P_4(z)}{\prod_{i=0}^2{(z-q_i)^2}},\quad P_4(z)=p_4^{(0)}(z)+\dfrac14(q-1)E\prod_{i=0}^2{(z-q_i)}.
\end{equation}
In \cite{Eguchi2010,Eguchi2010a}, this curve has been identified with the Seiberg-Witten curve \cite{Seiberg1994,*Seiberg1994a} of the $SU(2)$ superconformal gauge theory on $\mathbb{R}^4$ in the Gaiotto form \cite{Gaiotto2009a}. It describes a torus with singular points at $z=q_f$ which is a double cover of the four punctured Riemann sphere. In the complex $z$-plane, the function $H_0(z)$ has four branch points denoted $z_1$ to $z_4$,
\begin{equation}
P_4(z)=\dfrac{m_\infty^2}{4}\prod_{\a=1}^4{(z-z_\a)}.
\end{equation}
These branch points are functions of the masses $m_{f\in\{0,1,2,\infty\}}$, the Coulomb branch modulus $E$, and the gauge coupling $q$. The two dual cycles $\CA$ and $\CB$ on the torus will be defined as in the figure \refOld{fig1}.

In \cite{Eguchi2010a}, the Seiberg-Witten differential has been identified with $dS_0=2H_0(z)dz$, and the filling fraction condition provides the first Seiberg-Witten relation \cite{Seiberg1994,*Seiberg1994a},
\begin{equation}\label{SW_cl}
 a(E)=\oint_\CA\dfrac{dS_0}{2i\pi}, \qquad \dfrac{\p\CF_\text{SW}^{(0)}}{\p a}(E)=\oint_\CB{dS_0}.
\end{equation}
The planar free energy of the hermitian matrix model ($\b=1$) was shown to obey the second Seiberg-Witten relation \ref{SW_cl} in \cite{Chekhov2002}. It has been successfully compared to the limit $\CF_\text{SW}^{(0)}=\lim_{\e_1\to0}\CF_\text{SW}$ of the Nekrasov partition at the point $m_0=m_\infty=0$, $m_1=m_2$ \cite{Eguchi2010}. However, instead of performing the $\CB$-cycle integral, the $q$-derivative identity \ref{deriv_q} has been used, leaving the possibility to add a $q$-independent constant to the prepotential. The classical prepotential $\CF_\text{SW}^{(0)}$ can be recovered upon integration of the second Seiberg-Witten relation \ref{SW_cl}, after inversion of the first relation into $E(a)$.

\subsection{WKB expansion}\label{sec_WKB}
Pluging the WKB expansion \ref{WKB} in the Riccati equation \ref{equa_diff}, we obtain the recursion relation
\begin{equation}\label{rec_sol}
2H_0H_n=T_n-\p H_{n-1}-\sum_{k=1}^{n-1}{H_kH_{n-k}},
\end{equation}
where $T_n$ vanishes for $n>2$. This equation has been solved up to $n=2$, and used to recover the free energy at the order $O(\hbar^2)$ \cite{Nishinaka2011,Maruyoshi2010,Nishinaka2012}. It is possible to go further by using the ansatz $H_n=H_n^{(r)}+H_n^{(s)}H_0$ with two meromorphic functions $H_n^{(r)}$ and $H_n^{(s)}$. The regular term of this decomposition arises naturally from the Riccati equation and can be observed at the first orders. The second term assumes that the function has the same branch points as the $\hbar$-classical current $H_0(z)$. This is in the spirit of \cite{Mironov2010} where the quantum prepotential is obtain from the cycle integral of a deformed differential form, the underlying curve remaining classical. The opportunity to keep the classical spectral curve is due to the possibility of expanding in $\hbar$ the branch points $z_\a(\hbar)$ of a $\b$-deformed curve, $\sqrt{z-z_\a(\hbar)}$ being at each order $O(\hbar^n)$ a meromorphic function with a pole of degree $n$ at $z=z_\a(0)$ times the ``classical'' square root $\sqrt{z-z_\a(0)}$. In this setting, branch points of negative order arise naturally. We should stress that these poles of $H_n^{(r)}$ and $H_n^{(s)}$ are not related to the singularities of the Penner type potential \ref{Penner} but are a general property of the WKB expansion.

The recursion relation \ref{rec_sol} splits into two relations for regular and singular parts respectively,
\begin{align}
\begin{split}\label{rec_Hn}
&2H_n^{(r)}=-\p H_{n-1}^{(s)}-2\sum_{k=2}^{n-1}{H_k^{(r)}H_{n-k}^{(s)}},\\
&2H_n^{(s)}=\dfrac{T_n}{T_0}-\sum_{k=1}^{n-1}{H_k^{(s)}H_{n-k}^{(s)}}-\dfrac1{H_0}\p\left(\dfrac{H_{n-1}^{(r)}}{H_0}\right)-\dfrac1{H_0^2}\sum_{k=2}^{n-2}{H_k^{(r)}H_{n-k}^{(r)}},
\end{split}
\end{align}
where $n>2$ and we used the expression of $H_1^{(r)}=-(1/2)\p H_0/H_0$.

We now analyze the singularities of the functions $H_n^{(r)}$ and $H_n^{(s)}$, we expect poles located at the branch points $z=z_\a$ and at the singularities $z=q_i$ of the spectral curve. We first notice that the ratios $T_n/T_0$ have only simple poles at $z=z_\a$, and no poles at $z=q_i$, they behave at most as a constant when $z\to\infty$. Next, we look at the first terms of the recursion \ref{rec_Hn},
\begin{equation}
H_1^{(r)}(z)=\hf\sum_{i=0}^2\dfrac1{z-q_i}-\dfrac14\sum_{\a=1}^4\dfrac1{z-z_\a},\quad H_1^{(s)}(z)=\hf\dfrac{T_1(z)}{T_0(z)}.
\end{equation}
These two functions have simple poles at the branch points $z=z_\a$, and $H_1^{(r)}$ has additional poles at $z=q_i$. At infinity, $H_1^{(r)}$ behaves as $1/2z$ and $H_1^{(s)}$ as $O(1/z)$.

From the recursion relation \ref{rec_Hn}, it is possible to show that the functions $H_n^{(r)}$ and $H_n^{(s)}$ with $n\geq2$ have no poles at $z=q_i$. We further see that $H_n^{(r)}$ is at most $O(1/z^2)$ at infinity, and $H_n^{(s)}$ at most constant. These constants are actually vanishing due to the constraint $W(z)\sim \sqrt{\b}gN/z$ at $z\to\infty$ which imposes $H_n(z)\sim O(1/z^2)$ for $n\geq2$, and consequently $H_{n\geq1}^{(s)}(z)\sim O(1/z)$. Finally, we denote $\a_n^{(r)}$ and $\a_n^{(s)}$ the order of the pole at $z=z_\a$ for the functions $H_n^{(r)}$ and $H_n^{(s)}$. The recursion \ref{rec_Hn} suggests $\a_n^{(r)}=\a_{n-1}^{(s)}+1$ and $\a_n^{(s)}=\a_{n-1}^{(r)}+2$ which, together with the initial value $\a_1^{(r)}=\a_1^{(s)}=1$, is solved by  $\a_n^{(r)}=\lfloor(3n-1)/2\rfloor$ and $\a_n^{(s)}=\lfloor3n/2\rfloor$ where $\lfloor x\rfloor$ denotes the lower integer bound of $x$. Coming back to the two equations \ref{rec_Hn}, we check that indeed the orders of the poles at $z=z_\a$ in each terms of the LHS have upper bound $\a_n^{(r)}$ and $\a_n^{(s)}$ respectively. These analytical properties allow to write the functions $H_n^{(r)}$ and $H_n^{(s)}$ as
\begin{equation}\label{Hn}
H_n^{(r)}(z)=\sum_{p=1}^{\a_n^{(r)}}\sum_{\a=1}^4{\mu_{p,\a}^{(n)}(z-z_\a)^{-p}},\quad H_n^{(s)}(z)=\sum_{p=1}^{\a_n^{(s)}}\sum_{\a=1}^4{\mut_{p,\a}^{(n)}(z-z_\a)^{-p}},
\end{equation}
with moments $\mu$ and $\mut$ independent of $z$. These expressions are convenient to compute the $\CA$ and $\CB$ cycles integrals of $H(z)$ which can now be formulated as hypergeometric functions. They also provide a good starting point for the star operator construction in the CFT methods developed in \cite{Kostov1999,*Kostov2009,*Kostov2010} since they exhibit the singular expansion of the classical current $H(z)$ at the branch points. The moments $\mu$ and $\mut$ obey a recursion relation inherited from \ref{rec_Hn},
\begin{align}
\begin{split}\label{rec_mu}
\mu_{p,\a}^{(n)}&=\dfrac{p-1}{2}\mut_{p-1,\a}^{(n-1)}-\sum_{k=2}^{n-1}\Bigg(\sum_{q=1}^{\a_k^{(r)}}\sum_{r=1}^{\a_{n-k}^{(s)}}\mu_{q,\a}^{(k)}\mut_{r,\a}^{(n-k)}\d_{p-q-r}+\sum_{q=p}^{\a_k^{(r)}}\sum_{r=1}^{\a_{n-k}^{(s)}}\sum_{\b\neq\a}(-1)^{p-q}C_{q+r-p-1}^{r-1}z_{\a\b}^{p-q-r}\mu_{q,\a}^{(k)}\mut_{r,\b}^{(n-k)}\\
&+\sum_{q=1}^{\a_k^{(r)}}\sum_{r=p}^{\a_{n-k}^{(s)}}\sum_{\b\neq\a}(-1)^{p-r}C_{q+r-p-1}^{q-1}z_{\a\b}^{p-q-r}\mu_{q,\b}^{(k)}\mut_{r,\a}^{(n-k)}\Bigg)\\
\end{split}
\end{align}
and,
\begin{align}
\begin{split}\label{rec_mut}
2\mut_{p,\a}^{(n)}&=\d_{p,1}\dfrac{4p_4^{(n)}(z_\a)}{m_\infty^2\prod_{\b\neq \a}z_{\a\b}}-r_\a^{(0)}\sum_{k=1}^{n-1}\sum_{q=1}^{\a_k^{(r)}}{\sum_{r=1}^{\a_{n-k}^{(r)}}{\sum_{\superp{\b\neq\a}{\g\neq\a}}{\mu_{q,\b}^{(k)}\mu_{r,\g}^{(n-k)}z_{\a\b}^{-q}z_{\b\g}^{-r}}}}\\
&-\sum_{k=1}^{n-1}\Bigg(\sum_{q=1}^{\a_k^{(s)}}\sum_{r=1}^{\a_{n-k}^{(s)}}\mut_{q,\a}^{(k)}\mut_{r,\a}^{(n-k)}\d_{p-q-r}+2\sum_{q=p}^{\a_k^{(s)}}\sum_{r=1}^{\a_{n-k}^{(s)}}\sum_{\b\neq\a}(-1)^{p-q}C_{q+r-p-1}^{r-1}z_{\a\b}^{p-q-r}\mut_{q,\a}^{(k)}\mut_{r,\b}^{(n-k)}\Bigg)\\
&-\sum_{q=p-1}^{\a_{n-1}^{(r)}}{r_\a^{(q+1-p)}\mu_{q,\a}^{(n-1)}}-\d_{p,1}\sum_{q=1}^{\a_{n-1}^{(r)}}\sum_{\b\neq\a}r_\a^{(0)}\mu_{q,\b}^{(n-1)}z_{\a\b}^{-q}-\sum_{k=1}^{n-1}\sum_{m=0}^{\a_k^{(r)}+\a_{n-k}^{(r)}}\equskip r_\a^{(m)}\Bigg(\sum_{q=1}^{\a_k^{(r)}}\sum_{r=1}^{\a_{n-k}^{(r)}}\mu_{q,\a}^{(k)}\mu_{r,\a}^{(n-k)}\d_{p+m-q-r-1}\\
&-2\sum_{q=p+m-1}^{\a_k^{(r)}}\sum_{r=1}^{\a_{n-k}^{(r)}}\sum_{\b\neq\a}(-1)^{p+m-q}C_{q+r-p-m}^{r-1}z_{\a\b}^{p+m-q-r-1}\mu_{q,\a}^{(k)}\mu_{r,\b}^{(n-k)}\Bigg)
\end{split}
\end{align}
where we denoted $z_{\a\b}=z_\a-z_\b$, the binomial coefficient $C_p^q$, and
\begin{equation}
r_\a^{(n)}=\oint_{z_\a}{\dfrac{(z-z_\a)^{-n}}{H_0(z)^2}\dfrac{dz}{2i\pi}},\quad H_0(z)^{-2}=\sum_{n=0}^{\infty}{(z-z_\a)^{n-1}r_\a^{(n)}}.
\end{equation}

\subsection{Formal resummation}
The previous $\hbar$-expansion for singular and regular parts of $H$ can be formally resummed into the functions $H^{(r)}$ and $H^{(s)}$. These two functions are related to each other through
\begin{equation}
H^{(r)}(z)=-(\hbar/2)\p\log(H^{(s)}(z)H_0(z)),
\end{equation}
which is a resummed version of the first recursion relation in \ref{rec_Hn}. We further introduce the function $\eta(z)$ defined up to an additive constant by $H^{(s)}H_0=\p\eta$, so that $H^{(r)}=-(\hbar/2)\p\log\p \eta$. Reassembling singular and regular parts, we write the relation between $\eta$ and $H$ as
\begin{equation}\label{decompo_H}
H(z)=\p \eta(z)-\dfrac{\hbar}2\p\log\p\eta(z).
\end{equation}
Inserting this definition into the Riccati equation, we find an equation over $\eta$ which takes the form 
\begin{equation}\label{equ_h}
T(z)=(\p \eta(z))^2-\dfrac{\hbar^2}{2}\{\eta(z),z\},
\end{equation}
where $\{\eta,z\}$ denotes the Schwartz derivative. This expression arises naturally in the Schr\"odinger equation context, \footnote{We thank V. Pasquier for suggesting this connection.} and appears in the accessory parameters problem (see for instance \cite{Ferrari2012}). Indeed, combining the relation between $h$, $H$ and $\psi$, and given that $\eta$ and $-\eta$ satisfy the same equation \ref{equ_h}, the two independent solutions of the Schr\"odinger equation write
\begin{equation}\label{sol_Schrod}
\hbar^2\p^2\psi^\pm(z)=T(z)\psi^\pm(z) \implies \psi^\pm(z)=\sqrt{\dfrac{\hbar}2}\dfrac{e^{\pm\hbar^{-1}\eta(z)}}{\sqrt{\p \eta(z)}},
\end{equation}
where the constant factors are fixed by taking the Wronskian $\psi^-\p\psi^+-\psi^+\p\psi^-$ equal to one. At the order $O(\hbar)$, $\p \eta$ is equal to the classical momentum. The ratio of the two solutions, $r=\psi^-/\psi^+$ is known to satisfy $T=-(\hbar^2/2)\{r,z\}$, replacing $r=e^{-2\hbar^{-1}\eta}$ we recover the equation \ref{equ_h}.

There is another interesting interpretation for the expressions \ref{decompo_H} and \ref{equ_h}. In the CFT formulation of matrix models \cite{Marshakov1991,Kostov1999} extended to \bens s \cite{Dijkgraaf2009,Itoyama2009}, the partition function $\CZ_\b$ reproduces the correlator of a free field with a background charge proportional to $\hbar$. Up to a Wick rotation, this field is the holomorphic part of the Liouville field after integration over the zero mode. In this framework, the planar limit is the classical limit of this Coulomb gas field, and $H(z)$ is interpreted as the $g$-classical $U(1)$ current $H=\p\phi$. Similarly, $T(z)$ corresponds to the $g$-classical stress-energy tensor, and the Riccati equation \ref{equa_diff} simply express the standard definition $T=(\p\phi)^2+\hbar\p^2\phi$. The relation \ref{equ_h} between $T(z)$ and $\eta(z)$ can be seen as the transformation of a constant stress-energy tensor under the inverse mapping $\eta\to z(\eta)$.\footnote{In the context of Hitchin systems, this is the transformation law of a projective connection \cite{Teschner2012}.} When the background charge is non-zero, the $U(1)$ current is no longer a primary field, and the relation \ref{decompo_H} is the transformation of a constant current under the same conformal mapping $\eta\to z(\eta)$. Thus, in the coordinate $\eta$, both classical stress-energy tensor and $U(1)$ current are equal to one. Solving the Schr\"odinger equation in this coordinate, we get
\begin{equation}\label{coord_h}
\hbar^2\dfrac{\p^2}{\p \eta^2}\psi(\eta)=\psi(\eta)\implies \psi^\pm(\eta)=\sqrt{\dfrac{\hbar}2}e^{\pm\hbar^{-1}\eta},
\end{equation}
which indeed gives $H(\eta)=\hbar\p_\eta\log\psi^+(\eta)=1$. The transformation law of the wave functions $\psi^\pm$ can be deduced from their relation to the classical current, they transform as primary fields of dimension $-1/2$ under $\eta\to z(\eta)$,
\begin{equation}
\psi^\pm(\eta)\to (\p \eta(z))^{-1/2}\psi^\pm(\eta(z)),
\end{equation}
and we recover the expression \ref{sol_Schrod} from \ref{coord_h}.

\section{Deformed Seiberg-Witten theory}
\subsection{Planar free energy and Seiberg-Witten relations}\label{subsec_SW}
We now show that the planar free energy $\CF_\b$ of the \bens\ obeys the Seiberg-Witten relations with a deformed differential form $dS=2\p\eta dz$ given by only the singular part of the resolvent,
\begin{equation}\label{SW}
a(E)=\oint_\CA{\dfrac{dS}{2i\pi}},\qquad \dfrac{\p\CF_\b}{\p a}(E)=\dfrac14\oint_\CB{dS},
\end{equation}
with the cycles $\CA$ and $\CB$ previously defined on the classical spectral curve. Let us emphasize that we consider here the formal resumation in $\hbar$, and these relations, together with all the identities below, only hold order by order in the $\hbar$ expansion. In the limit $\hbar\to0$, $\p \eta$ reduces to $H_0$ and we recover the standard Seiberg-Witten relations \ref{SW_cl}.

To analyze the first relation in \ref{SW}, we decompose $H$ into regular and singular parts as in \ref{decompo_H},
\begin{equation}\label{int_H}
\oint_\CA{H(z)dz}=\oint_\CA{\p\eta dz}-\dfrac{\hbar}{2}\oint_\CA{\dfrac{\p H_0(z)}{H_0(z)}dz}-\dfrac{\hbar}{2}\oint_\CA{\p\log(H^{(s)}(z))dz}.
\end{equation}
The last term in the RHS is proportional to the monodromy of $\log(H^{(s)}(z))$ around the cycle $\CA$. Since the expansion of the logarithm in $\hbar$ produces only regular terms at each order, it is vanishing. The second term in the RHS of \ref{int_H} exactly compensates the shift of $-\hbar$ for the filling fraction condition observed in \cite{Nishinaka2011},
\begin{equation}\label{filling_frac}
\oint_\CA{2H(z)\dfrac{dz}{2i\pi}}=a-\hbar.
\end{equation}
This relation is believed to be exact at the order $O(\hbar^2)$ \cite{Nishinaka2012}.\\

To demonstrate the second relation \ref{SW}, we employ the collective field formulation of the \bens\ at large $N$ derived by Dyson \cite{Dyson1962}. A similar calculation has been done in \cite{Chekhov2002} for $\hbar=0$ and we extend it here to the $\b$-deformation. Closely related considerations are found in the study of complex \bens s in which the eigenvalues are integrated over the full complex plane, leading to a bidimensional Dyson gas \cite{Wiegmann2005}.

We first review the derivation of the Dyson action \cite{Dyson1962}. In the large $N$ limit, the \bens\ partition function can be written as a path integral over the density of eigenvalues $\rho(x)$ seen as a collective field,
\begin{equation}\label{collec_field}
\CZ_\b=\int{D[\rho]e^{\frac1{g^2}\CA[\rho]}},\quad \rho(x)=g\sqrt{\b}\la\sum_{I=1}^N\d(x-\l_I)\ra.
\end{equation}
Since we work perturbatively in $\hbar$, the density at each order has a finite support $\G$ corresponding to the branch cuts of the classical resolvent $H_0(z)$. Again, this is no longer true at finite $\hbar$ where the eigenvalues remain isolated. The action appearing in \ref{collec_field} is obtained from the effective action for the eigenvalues $\l_I$ deduced from the definition \ref{def_beta_ens},
\begin{equation}
\dfrac1{g^2}\CA_\text{eff}[\l_I]=\dfrac{\sqrt{\b}}{g}\sum_{I=1}^N{V(\l_I)}+\b\sum_{I,J=1}^N{\log|\l_I-\l_J|}-\b\sum_{I=1}^N{\log l(\l_I)},
\end{equation}
where we introduced a short distance cut-off $l(x)$ to regularize the sum of logarithms at coincident indices $I=J$. This cut-off is interpreted as the mean distance between the eigenvalues at the point $x$, it also arises in the Jacobian of the transformation,
\begin{equation}
\prod_{I=1}^N{d\l_I}=N! J[\rho]D[\rho],\quad J[\rho]\propto\prod_{I=1}^N{l(\l_I)}.
\end{equation}
Combining these two contributions, and introducing the density within the effective action, we derive the following collective field action 
\begin{equation}
\CA[\rho]=\int{V(x)\rho(x)dx}+\int{\log|x-y|\rho(x)\rho(y)dxdy}-\hbar\int{\rho(x)\log l(x)dx}+\text{cst}.
\end{equation}
Since a ``single'' eigenvalue roughly occupies a segment of size proportional to the inverse of the local density, $l(x)\propto1/\rho(x)$ and the classical action reads
\begin{equation}\label{action_eff}
\CA[\rho]=\int_\G{V(x)\rho(x)dx}+\int_\G{\log|x-y|\rho(x)\rho(y)dxdy}+\hbar\int_\G{\rho(x)\log \rho(x)dx},
\end{equation}
up to a constant set to zero by renormalization of $\CZ_\b$.

To confirm the expression \ref{action_eff} of the collective field action, we should analyze the equation of motion,
\begin{equation}\label{eom}
\dfrac{\p}{\p x}\dfrac{\d\CA[\rho]}{\d\rho(x)}=0\implies V'(x)+2\int_\G{\dfrac{\rho(y)dy}{x-y}}+\hbar\dfrac{\rho'(x)}{\rho(x)}=0,\quad x\in\G,
\end{equation}
where the integral is considered as a principal value. The density solving this equation of motion is the $g$-classical, or planar, density, i.e. the limit of the density $\rho(x)$ defined in \ref{collec_field} as $g\to0$. Since we work only at the planar level, and the full density $\rho(x)$ will play no role below, we will keep the notation $\rho(x)$ now referring to the planar density. It is related to the planar resolvent \ref{def_resol} by a Cauchy transform
\begin{equation}
W(z)=\int_\G{\dfrac{\rho(y)dy}{z-y}},\quad W(x\pm i0)=\mp i\pi\rho(x)+\int_\G{\dfrac{\rho(y)dy}{x-y}},\quad x\in\G.
\end{equation}
Taking the discontinuity of the loop equation \ref{le1} over the branch cut $\G$, we recover the equation of motion \ref{eom}. In particular, the derivative of the resolvent which characterizes the $\b$-deformation is at the origin of the logarithmic derivative term $\p\log\rho$ in \ref{eom}.

The classical spectral curve \ref{spec_curve} describes a two cuts solution: the density support $\G$ is made of two disjoint intervals $\G=\G_-\cup\G_+$ with $\G_-=[z_1,z_2]$ and $\G_+=[z_3,z_4]$. The moduli of the solution are fixed by imposing the two filling fraction conditions
\begin{equation}\label{cond_mod}
\int_{\G_-}{\rho(x)dx}=\hf(a-\hbar),\quad \int_{\G_+}{\rho(x)dx}=\sqrt{\b}gN-\hf\left(a-\hbar\right).
\end{equation}
The first identity is equivalent to the Seiberg-Witten relation \ref{SW} involving the $\CA$-cycle integral which circles the branch cut $\G_-$. The second relation reflects the normalization of the density to $\sqrt{\b}gN$.

The planar free energy is the collective field action evaluated for the planar density that solves the equation of motion \ref{eom}. Taking the $a$-derivative of the expression \ref{action_eff}, we get
\begin{equation}
\dfrac{\p \CF_\b}{\p a}=\int_\G{\dfrac{\p\rho(x)}{\p a}C(x)},\quad\text{with } C(x)=V(x)+2\int_\G{\log|x-y|\rho(y)dy}+\hbar\log\rho(x).
\end{equation}
The equation of motion \ref{eom} implies that $C(x)$ is a constant on each support $\G_\pm$ that we denote $C_\pm$. In the standard approach to matrix models, these two constants are supposed to be equal,which provides an extra requirement on $\rho$ \cite{Jurkiewicz1990}. As mentioned in \cite{Chekhov2002}, this is no longer the case in the Seiberg-Witten theory. The derivative of the filling fraction conditions \ref{cond_mod} with respect to $a$ supplies two expressions that can be used to rewrite the free energy derivative as the difference of the two constants.
\begin{equation}
\dfrac{\p \CF_\b}{\p a}=\hf(C_--C_+).
\end{equation}
To rewrite this equation as a cycle integral, we need to extend the function $C(x)$ to the whole complex plane, i.e. we are looking for a function $C(z)$ with a branch cut on $\G$ where it satisfies $C(x+i0)+C(x-i0)=2C(x)$. This function can be chosen up to a constant term since we are only interested in the difference $C_--C_+$. Once it has been properly defined, it is possible to write 
\begin{equation}\label{SW_rel2}
\dfrac{\p \CF_\b}{\p a}=-\hf\int_{z_2}^{z_3}{\dfrac{dC}{dz}dz}=\dfrac14\oint_\CB{\p C(z)dz},
\end{equation}
provided that $\p C$ has no regular part. When $\b=1$, the function $C(z)$ is the classical bosonic field appearing in the CFT methods  \cite{Marshakov1991,Kostov1999,Dijkgraaf2009,Itoyama2009}, $C(z)=2\phi(z)$ with $H(z)=\p\phi(z)$. When $\hbar\neq0$, we need an additional contribution $C(z)=2\phi(z)+\hbar L(z)$ because of the logarithmic term. This function $L(z)$ have a branch cut on $\G$ where it verifies
\begin{equation}
L(x+i0)+L(x-i0)=2\log\rho(x).
\end{equation}
To identify the function $L(z)$, we notice that since $\p \eta$ is by definition the singular part of $H$, it satisfies $\p \eta(x\pm i0)=\mp i\pi\rho(x)$ for $x\in\G$. It implies
\begin{equation}
\log(\p \eta(x+i0))+\log(\p \eta(x-i0))=\log(\pi^2\rho(x)^2),
\end{equation}
which leads to identify $L$ with $\log\p \eta$ up to an irrelevant additive constant. We conclude that $C(z)=2\eta(z)$, and \ref{SW_rel2} provides the second Seiberg-Witten relation \ref{SW}. The difference $H^{(r)}=(\hbar/2)\p L$ between the two functions $H(z)$ and $\p \eta(z)$ is specific to the $\b$-deformation. It is related to the cut-off part of the Dyson action, i.e. the term $\rho\log\rho$ in \ref{action_eff}. The $\CB$-cycle integral of the relation \ref{SW} is of course invariant under a regular shift of the differential, but again $H^{(r)}$ has poles at the branch points which require a special treatment.\\

To summarize, we have shown that the Dyson representation of the planar free energy,
\begin{equation}\label{F_b}
\CF_\b=\oint_\G{\dfrac{dz}{2i\pi}\left[V(z)+\hbar\log\p\eta(z)\right]\p \eta(z)\ }+\oint_{\G}{\dfrac{dz}{2i\pi}\oint_\G{\dfrac{dz'}{2i\pi}\log|z-z'|\p \eta(z)\p \eta(z')}},
\end{equation}
provides a new solution to the Seiberg-Witten relations \ref{SW}. The potential $V(z)$ is chosen such that the function $T(z)$ defined in \ref{def_HT} satisfies the equation \ref{equ_h} in which the RHS is entirely defined by the differential form $dS=2\p\eta dz$. In \ref{F_b}, the contours circle both branch cuts $\G_-$ and $\G_+$ but exclude the logarithmic branch points of the potential. In addition, the double integral in the RHS must be properly regularized at coincident point $z=z'$.

\subsection{Differential operator}
In their study of the deformed Seiberg-Witten relations for the pure $SU(2)$ $\mathcal{N}=2$ gauge theory, Mironov and Morozov \cite{Mironov2010} introduced a differential operator in the Coulomb branch modulus and the gauge coupling to derive the $O(\hbar^2)$ corrections to the quantized differential form $dS$. A similar operator with derivatives in the parameters of the potential was observed in models with Gaussian and cubic potentials \cite{Aganagic2011}. In $SU(2)$ superconformal QCD, such a differential operator also appeared at first order in $O(\hbar)$ \cite{Maruyoshi2010}, with a differentiation with respect to the masses. Here we demonstrate the existence of such operators at all order in $\hbar$. These relations ensures that the quantities considered in the last subsection are finite.

The summation of fractions in the expression \ref{Hn} of $H_n^{(s)}$ can be reorganized as
\begin{equation}
H_n^{(s)}(z)=\sum_{p=1}^{\a_n^{(s)}}{\dfrac{a_pz^3+b_pz^2+c_pz+d_p}{P_4(z)^p}}.
\end{equation}
A recursion relation on the numerators can be derived directly from \ref{rec_Hn}, and would involve Euclidean divisions of polynomials. Here we do not need the explicit expression of the coefficients $a_p,\ b_p,\ c_p$ and $d_p$ which are functions of the masses, the Coulomb branch modulus, and the gauge coupling. The degree four polynomial $P_4(z)$ decomposes as a sum of monomials,
\begin{equation}
P_4(z)=\dfrac{m_\infty^2}{4}z^4+f_3z^3+f_2z^2+f_1z+f_0,
\end{equation}
where the coefficients $f_j$ depends on the four masses $m_f$, as well as the parameters $q$ and $E$. Taking the derivative of the spectral curve \ref{spec_curve} with respect to $f_j$, we easily show that
\begin{equation}
\dfrac{\p}{\p f_j}\left(\dfrac{\p}{\p f_0}\right)^{p-1}H_0(z)=(-1)^{p-1}\dfrac{\G[p-1/2]}{2\G[1/2]}\dfrac{z^{j}}{P_4(z)^p}H_0(z).
\end{equation}
Using this identity, we rewrite the $n$-th term in the $\hbar$-expansion of the differential $\p\eta$ as a differential operator in $f_j$ acting on $H_0$,
\begin{equation}
H_n^{(s)}H_0=\left[\sum_{p=1}^{\a_n^{(s)}}{(-1)^{p-1}\dfrac{2\G[1/2]}{\G[p-1/2]}\left(\dfrac{\p}{\p f_0}\right)^{p-1}\left(a_p\dfrac{\p}{\p f_3}+b_p\dfrac{\p}{\p f_2}+c_p\dfrac{\p}{\p f_1}+d_p\dfrac{\p}{\p f_0}\right)}\right]H_0
\end{equation}
It now remains to change variable from $f_j$ to the three masses $m_0,m_1,m_2$ and the Coulomb branch modulus $E$. The Jacobian matrix of this change of variables is given by
\begin{equation}\label{Jacobian}
 \left (
   \begin{array}{c}
      \frac{\p}{\p f_3} \\
      \frac{\p}{\p f_2} \\
      \frac{\p}{\p f_1} \\
      \frac{\p}{\p f_0} 
   \end{array}
   \right )=
\begin{pmatrix}\frac{2\,q}{m_2\,{\left( q-1\right) }^{2}} & \frac{2}{m_1\,{\left( q-1\right) }^{2}} & 0 & -\frac{8\,{q}^{2}}{{\left( q-1\right) }^{3}}\cr \frac{2}{m_2\,{\left( q-1\right) }^{2}} & \frac{2}{m_1\,{\left( q-1\right) }^{2}} & 0 & -\frac{4\,\left( 3\,q-1\right) }{{\left( q-1\right) }^{3}}\cr \frac{2}{m_2\,{\left( q-1\right) }^{2}\,q} & \frac{2}{m_1\,{\left( q-1\right) }^{2}} & 0 & -\frac{8\,\left( 2\,q-1\right) }{{\left( q-1\right) }^{3}\,q}\cr \frac{2}{m_2\,{\left( q-1\right) }^{2}\,{q}^{2}} & \frac{2}{m_1\,{\left( q-1\right) }^{2}} & \frac{2}{m_0\,{q}^{2}} & -\frac{4\,\left( {q}^{3}-3\,{q}^{2}+8\,q-4\right) }{{\left( q-1\right) }^{3}\,{q}^{2}}\end{pmatrix}
\left (
   \begin{array}{c}
      \frac{\p}{\p m_2} \\
      \frac{\p}{\p m_1} \\
      \frac{\p}{\p m_0} \\
      \frac{\p}{\p E} 
   \end{array}
   \right )
\end{equation}
Its determinant is non-vanishing as long as the four punctures $q_f$ of the Riemann sphere remain separated, i.e. $q\neq0$ and $q\neq 1$. At first order in $\hbar$, we recover the result found in \cite{Maruyoshi2010},
\begin{equation}\label{Op_1}
H_1^{(s)}H_0=-\left(\dfrac{\p}{\p m_1}+\dfrac{\p}{\p m_2}\right) H_0.
\end{equation}
This differential operator does not involve the derivative with respect to the Coulomb parameter $E$. This simplification is due to the previous change of variable \ref{def_E} from $u$ to $E$. We would have obtained exactly the same property if we had chosen instead of $E$ the variable $q\p\CF_\text{SW}/\p q$ related to $u$ through \ref{deriv_q}. However, the intermediate formulas such as the expression of the polynomials \ref{expr_poly}, or the Jacobian matrix \ref{Jacobian}, would have been more complicated. It is not clear priori whether the higher order operators would involve a differentiation with respect to $E$, or if it can again be eliminated after a redefinition of the Coulomb branch parameter.

\subsection{Interpretation of the differential form in Seiberg-Witten theory}
We have shown in subsection \refOld{subsec_SW} that the planar free energy obeys the Seiberg-Witten relation with the differential $dS=2\p\eta dz$. It remains to see whether this differential is relevant to the Seiberg-Witten theory. In particular, since the Seiberg-Witten differential is commonly identified with the full resolvent $2H(z)dz$, we should compare here $H$ with $\p\eta$. We have seen previously that the $\CA$-cycle integral reproduces the correct result, including the shift of $-\hbar$ observed in \cite{Nishinaka2011}. The Bohr-Sommerfeld quantization rule applied to the underlying quantum integrable model provides a simple argument in favor of replacing $H$ by $\p\eta$: the classical momentum appearing within the cycle integrals is equal to $\p \eta(z)$ up to terms of order $O(\hbar^2)$, and to $H(z)$ only up to $O(\hbar)$. In this context, the shift of $-\hbar$ has indeed to be taken into account.

We now focus on the $\CB$-cycle integral. As a consequence of the relation \ref{Op_1}, the planar free energy obeys the following identity,
\begin{equation}\label{CF_b_1}
\dfrac{\p}{\p a}\CF_\b^{(1)}=-\left(\dfrac{\p}{\p m_1}+\dfrac{\p}{\p m_2}\right)\dfrac{\p}{\p a}\CF_\b^{(0)},
\end{equation}
where the superscript pertains to the $\hbar$-expansion of $\CF_\b$. The equality \ref{AGT_planar} between the prepotential and the free energy has been checked at first orders in $q$ up to a $q$-independent term in \cite{Nishinaka2011}. A priori, such a $q$-independent but $a$-dependent term could be related to a non-zero $\CB$-cycle integral of $H_1^{(r)}$, and would break the relation \ref{CF_b_1} for the Seiberg-Witten prepotential. In order to discard this term, we have to verify that the prepotential $\CF_\text{SW}$ at the order $O(q^0)$, i.e. the one-loop contribution, satisfies \ref{CF_b_1}.\footnote{This is trivial for the classical part $\CF_\text{SW,cl}=a^2\log q$.} The expression of the one-loop correction to the prepotential can be found in \cite{Alday2009},
\begin{align}
\begin{split}
&\CF_\text{SW,1-loop}=\CF_\text{vector}(a)+\sum_\pm\CF_\text{hyper}(\mu_i,\pm a),\quad \CF_\text{vector}(a)=\g_2(2a)+\g_2(2a+\e_1),\\
&\CF_\text{hyper}(\mu_i,a)=-\g_2(a-\mu_1+\e_1)-\g_2(a-\mu_2+\e_2)-\g_2(-a+\mu_3)-\g_2(-a+\mu_4),
\end{split}
\end{align}
where we separated the $\mathcal{N}=2$ vector and hypermultiplet contributions, and denoted $\g_2$ the limit $\e_2\to0$ of the logarithm of Barnes' double gamma function $\G_2(x|\e_1,\e_2)$,
\begin{equation}
\g_2(x)=\lim_{\e_2\to0}{\e_1\e_2\log\G_2(x|\e_1,\e_2)}.
\end{equation}
The $\e_1$-expansion of the function $\g_2$ satisfies the remarkable property
\begin{equation}
\g_2(x)=\left(1-\dfrac{\e_1}{2}\dfrac{d}{dx}+\dfrac{\e_1^2}{12}\dfrac{d^2}{dx^2}\right)\g(x)+O(\e_1^3),\quad \g(x)=\lim_{\e_1\to0}\g_2(x).
\end{equation}
Replacing the derivative $\g'(x)$ by a differentiation with respect to the $\mu_i$, and then changing variable from $\mu_i$ to $m_f$, it is possible to show that
\begin{equation}
\CF_\text{hyper}^{(1)}(\mu_i,a)=-\left.\left(\dfrac{\p}{\p m_1}+\dfrac{\p}{\p m_2}\right)\right|_\text{$a$ fixed}\CF_\text{hyper}^{(0)}(\mu_i,a).
\end{equation}
In this calculation we considered $a$ fixed, i.e. independent of the variables $\mu_i$ (or $m_f$) and $\hbar$. Since at the order $O(\e_1)$ the contribution of the vector multiplet is vanishing, the previous relation extends to the whole one-loop contribution,
\begin{equation}
\CF_\text{SW,1-loop}^{(1)}=-\left.\left(\dfrac{\p}{\p m_1}+\dfrac{\p}{\p m_2}\right)\right|_\text{$a$ fixed}\CF_\text{SW,1-loop}^{(0)}.
\end{equation}
We have not shown yet that the deformed Seiberg-Witten prepotential at the order $O(q^0)$ obeys the relation \ref{CF_b_1}. There is indeed an important subtlety since we worked with $a$ fixed, and the relation \ref{CF_b_1} holds only for $E$ fixed. It is actually possible to rewrite this relation at $a$ fixed, taking care of the reshuffling of the terms in the $\hbar$-expansion.\footnote{Let $a$ and $g$ be two functions of $E$, $\hbar$ and some parameter $m$. Changing variable from $E$ to $a$, the $\hbar$-expansion $g(E,\hbar)=g^{(0)}(E)+\hbar g^{(1)}(E)+O(\hbar^2)$ is reshuffled into $g(E(a,\hbar),\hbar)=\tilde{g}^{(0)}(a)+\hbar \tilde{g}^{(1)}(a)+O(\hbar^2)$. The relation $g^{(1)}(E)=\p_m|_E g^{(0)}(E)$ is equivalent to $\tilde{g}^{(1)}(a)=\p_m|_a \tilde{g}^{(0)}(a)$ provided that the second term in the $\hbar$-expansion of $a(E,\hbar)$ obeys $a^{(1)}=\p_m a^{(0)}$, since $g^{(0)}(E)=\tilde{g}^{(0)}(a^{(0)}(E))$ and $\p_m|_a=\p_m|_E-\p_ma^{(0)}(E)|_E\p_{a^{(0)}}$.} Then, this result dismisses the regular contribution $H_1^{(r)}$ and confirms the identification of $dS=2\p\eta dz$ with the Seiberg-Witten differential, up to terms of order $O(\hbar^2)$.

Another important feature of the Seiberg-Witten differential is the residues at the punctures. The classical differential $dS_0=2H_0(z)dz$ is known to have residues $m_f$ at the $z=q_f\in\{0,1,q,\infty\}$. Both $2H(z)dz$ and $2\eta(z)dz$ get $\hbar$-corrections that take the form of constant shifts: $m_0\to m_0+\hbar$ and $m_\infty\to m_\infty-\hbar$ for the first one, and $m_{1,2}\to m_{1,2}-\hbar$ for the later. Up to now, there is no interpretation for such a shift, but it is possible to correct the differentials by adding a pole contribution at $z=q_f$ that does not modify the $\CA$ and $\CB$ cycle integrals.

\section{Discussion: a direct approach to the prepotential}\label{sec_III}
In the Seiberg-Witten context, $E$ is seen as a parameter independent of $\hbar$, and $a$ is a function of $E$ and $\hbar$. This is rather unusual from the matrix model perspective where $E$ is a function of $a$, which is fixed, and $\hbar$. But both approaches are related: $E$ is an intermediate parameter of the Seiberg-Witten theory, it can be taken as any function $E(a_\text{fixed},\hbar)$. When this function is the inverse function of $a(E,\hbar)$ determined by the first Seiberg-Witten relation \ref{SW}, we recover the matrix model point of view.

It is instructive to take this later standpoint, and compose the WKB-expansion of $E$ and $H$, thus defining a new expansion of the classical current \cite{Maruyoshi2010},
\begin{equation}
E=\sum_{n=0}^\infty{\hbar^nE_n},\qquad H(z|E)=\sum_{n=0}^\infty{\hbar^n\tH_n(z|E_0,\cdots,E_n)},
\end{equation}
where we highlighted the dependence in $E$. The WKB method presented in the section \refOld{sec_WKB} still works in the same way, replacing $H_n$ by $\tH_n$, $E$ by $E_0$ and shifting
\begin{equation}
T_n\to \tilde{T}_n=T_n+\dfrac{(q-1)E_n}{4\prod_{i=0}^2{(z-q_i)}},\quad \text{for  }n\geq1,
\end{equation}
the quantities $\tilde{T}_n$ with $n>2$ being now non-vanishing. We notice that, $H_0(z|E)=\tH(z|E_0)$, and $\tH_n$ still decomposes into regular and singular parts which obey similar analytical properties. In particular, we may adapt the decomposition \ref{Hn} to $\tH_n$ and write
\begin{equation}\label{tHn}
\tH_n^{(r)}(z)=\sum_{p=1}^{\a_n^{(r)}}\sum_{\a=1}^4{\mu_{p,\a}^{(n)}(z-z_\a)^{-p}},\quad \tH_n^{(s)}(z)=\dfrac18\dfrac{(q-1)E_n}{P_4(z)}\prod_{i=0}^2{(z-q_i)}+\sum_{p=1}^{\a_n^{(s)}}\sum_{\a=1}^4{\mut_{p,\a}^{(n)}(z-z_\a)^{-p}}.
\end{equation}
Because of the shift in $\tH_n^{(s)}$, the coefficients $\mu_{p,\a}^{(n)}$ and $\mut_{p,\a}^{(n)}$ depend only on $E_0,\cdots,E_{n-1}$, they obey a slightly different recursion relation than \ref{rec_mu} and \ref{rec_mut} which is derived in a similar way. The Seiberg-Witten relations should also be expanded in $\hbar$, the zeroth order determines the relation between $a$ and $E_0$ as
\begin{equation}\label{SW_0}
\oint_\CA{2H_0(z|E_0)\dfrac{dz}{2i\pi}}=a,
\end{equation}
where now $a$ is a fixed parameter. The next orders give the dependence of $E_n(E_0,\cdots,E_{n-1})$ which is, up to a $q$ integration, the planar free energy of the \bens, as seen from the relation \ref{deriv_q},\footnote{We used here
\begin{equation}
\dfrac18(q-1)\oint_\CA{\dfrac{dz}{\sqrt{P_4(z)}}}=\dfrac{\p}{\p E_0}\oint_\CA{H_0(z|E_0)dz}.
\end{equation}}
\begin{equation}\label{SW_n}
E_n(E_0,\cdots,E_{n-1})=-\dfrac{dE_0}{da}\sum_{p=1}^{\a_n^{(s)}}\sum_{\a=1}^4{\mut_{p,\a}^{(n)}\oint_{\CA}\dfrac{H_0(z|E_0)}{(z-z_\a)^{p}}\dfrac{dz}{2i\pi}},
\end{equation}
This method produces directly the prepotential $\CF_\b$, up to terms independent of $q$. And again, a differential operator may be used to obtain the contour integral from \ref{SW_0}.

The previous calculation generates the $\b$-deformed electric prepotential $\CF_\b$, but it may also be interesting to determine directly the dual magnetic prepotential, related to $\CF_\b$ by a Legendre transform \cite{Galakhov2012}. Replacing the $\CA$-cycle by $\CB$-cycle integrals in the formulas \ref{SW_0} and \ref{SW_n}, we obtain $E_0(a_D)$ and $E_n(E_0,\cdots,E_{n-1})$. Finally, we may use the identity \ref{deriv_q} with $\CF_\b$ replaced by $\CF_\b^D$ and the $q$-derivative at fixed $a$ by a derivative at fixed $a_D$ to recover the dual prepotential upon integration. Such a formulation may help to understand the S-duality from the \bens\ point of view.


\section*{Acknowledgements}
This paper presents satellite results of a common project with Takahiro Nishinaka and Chaiho Rim, to whom I am indebted for numerous enlightening discussions and a careful reading of preliminary versions of this manuscript. I also would like to thank the hospitality of the CEA Saclay were part of this work was completed, and B. Eynard, I. Kostov, V. Pasquier and S. Ribault for very valuable discussions. This work is partially supported by the National Research Foundation of Korea (KNRF) grant funded by the Korea government (MEST) 2005-0049409.

\bibliographystyle{unsrt}

\begin{thebibliography}{10}

\bibitem{Nekrasov2009}
N.~Nekrasov and S.~Shatashvili.
\newblock {Quantization of Integrable Systems and Four Dimensional Gauge
  Theories}.
\newblock 2009.

\bibitem{Seiberg1994}
N.~Seiberg and E.~Witten.
\newblock {Monopole Condensation, And Confinement In N=2 Supersymmetric
  Yang-Mills Theory}.
\newblock {\em Nucl. Phys.}, B426:19--52, 1994.

\bibitem{Seiberg1994a}
N.~Seiberg and E.~Witten.
\newblock {Monopoles, duality and chiral symmetry breaking in N=2
  supersymmetric QCD}.
\newblock {\em Nucl. Phys.}, B431:484--550, 1994.

\bibitem{Gorsky1995}
A.~Gorsky, I.~Krichever, A.~Marshakov, A.~Mironov, and A.~Morozov.
\newblock {Integrability and Seiberg-Witten exact solution}.
\newblock {\em Phys. Lett.}, B355:466--474, 1995.

\bibitem{Yang1968}
Chen-Ning Yang and C.~P. Yang.
\newblock {Thermodynamics of a one-dimensional system of bosons with repulsive
  delta-function interaction}.
\newblock {\em J. Math. Phys.}, 10:1115--1122, 1969.

\bibitem{Nekrasov2003}
N.~Nekrasov.
\newblock {Seiberg-Witten prepotential from instanton counting}.
\newblock 2003.

\bibitem{Nekrasov2003a}
N.~Nekrasov and A.~Okounkov.
\newblock {Seiberg-Witten theory and random partitions}.
\newblock 2003.

\bibitem{Mironov2010}
A.~Mironov and A.Morozov.
\newblock {Nekrasov Functions and Exact Bohr-Sommerfeld Integrals}.
\newblock {\em JHEP}, 04:040, 2010.

\bibitem{Mironov2010a}
A.~Mironov and A.~Morozov.
\newblock {Nekrasov Functions from Exact BS Periods: the Case of SU(N)}.
\newblock {\em J.Phys.A}, 43:195401, 2010.

\bibitem{Popolitov2010}
A.~Popolitov.
\newblock {On relation between Nekrasov functions and BS periods in pure SU(N)
  case}.
\newblock 2010.

\bibitem{Alday2009}
L.~Alday, D.~Gaiotto, and Y.~Tachikawa.
\newblock {Liouville Correlation Functions from Four-dimensional Gauge
  Theories}.
\newblock {\em Lett. Math. Phys.}, 91:167--197, 2010.

\bibitem{Wyllard2009}
N.~Wyllard.
\newblock {A\_{N-1} conformal Toda field theory correlation functions from
  conformal N=2 SU(N) quiver gauge theories}.
\newblock {\em JHEP}, 11:002, 2009.

\bibitem{Bonelli2009}
G.~Bonelli and A.~Tanzini.
\newblock {Hitchin systems, N=2 gauge theories and W-gravity}.
\newblock {\em Phys. Lett.}, B691:111--115, 2010.

\bibitem{Mironov2010b}
A.~Mironov and A.~Morozov.
\newblock {On AGT relation in the case of U(3)}.
\newblock {\em Nucl.Phys.B}, 825:1--37, 2010.

\bibitem{Mironov2009a}
A.~Mironov and A.~Morozov.
\newblock {The Power of Nekrasov Functions}.
\newblock {\em Phys.Lett.B}, 680:188--194, 2009.

\bibitem{Giribet2009}
G.~Giribet.
\newblock {On triality in N=2 SCFT with N\_f=4}.
\newblock {\em JHEP}, 01:097, 2010.

\bibitem{Marshakov2009a}
A.~Marshakov, A.~Mironov, and A.~Morozov.
\newblock {Zamolodchikov asymptotic formula and instanton expansion in N=2 SUSY
  N\_f=2N\_c QCD}.
\newblock {\em JHEP}, 11:048, 2009.

\bibitem{Nanopoulos2009}
D.~Nanopoulos and Dan Xie.
\newblock {On Crossing Symmmetry and Modular Invariance in Conformal Field
  Theory and S Duality in Gauge Theory}.
\newblock {\em Phys. Rev.}, D80:105015, 2009.

\bibitem{Alba2009}
V.~Alba and A.~Morozov.
\newblock {Non-conformal limit of AGT relation from the 1-point torus conformal
  block}.
\newblock {\em JETP Lett.}, 90:708--712, 2009.

\bibitem{Marshakov2009b}
A.~Marshakov, A.~Mironov, and A.~Morozov.
\newblock {On Combinatorial Expansions of Conformal Blocks}.
\newblock {\em Theor. Math. Phys.}, 164:831--852, 2010.

\bibitem{Belavin2011}
A.~Belavin and V.~Belavin.
\newblock {AGT conjecture and Integrable structure of Conformal field theory
  for c=1}.
\newblock {\em Nucl. Phys.}, B850:199--213, 2011.

\bibitem{Kanno2011}
S.~Kanno, Y.~Matsuo, and S.~Shiba.
\newblock {W(1+infinity) algebra as a symmetry behind AGT relation}.
\newblock {\em Phys. Rev.}, D84:026007, 2011.

\bibitem{Mironov2009}
A.~Mironov and A.~Morozov.
\newblock {Proving AGT relations in the large-c limit}.
\newblock {\em Phys.Lett.B}, 682:118--124, 2009.

\bibitem{Fateev2009}
V.~A. Fateev and A.~V. Litvinov.
\newblock {On AGT conjecture}.
\newblock {\em JHEP}, 02:014, 2010.

\bibitem{Hadasz2010}
L.~Hadasz, Z.~Jaskolski, and P.~Suchanek.
\newblock {Proving the AGT relation for N\_f = 0,1,2 antifundamentals}.
\newblock {\em JHEP}, 06:046, 2010.

\bibitem{Marshakov1991}
A.~Marshakov, A.~Mironov, and A.~Morozov.
\newblock {Generalized matrix models as conformal field theories: Discrete
  case}.
\newblock {\em Phys. Lett.}, B265:99--107, 1991.

\bibitem{Kharchev1992}
S.~Kharchev, A.~Marshakov, A.~Mironov, A.~Morozov, and S.~Pakuliak.
\newblock {Conformal matrix models as an alternative to conventional
  multimatrix models}.
\newblock {\em Nucl. Phys.}, B404:717--750, 1993.

\bibitem{Kostov1999}
I.~Kostov.
\newblock {Conformal Field Theory Techniques in Random Matrix models}, 1999.
\newblock Based on the talk of the author at the Third Claude Itzykson Meeting,
  Paris, July 27-29, 1998.

\bibitem{Kostov2009}
I.~Kostov.
\newblock {Matrix models as conformal field theories: genus expansion}.
\newblock {\em Nucl. Phys.}, B837:221--238, 2010.

\bibitem{Kostov2010}
I.~Kostov and N.~Orantin.
\newblock {CFT and topological recursion}.
\newblock {\em JHEP}, 11:056, 2010.

\bibitem{Dijkgraaf2009}
R.~Dijkgraaf and C.~Vafa.
\newblock {Toda Theories, Matrix Models, Topological Strings, and N=2 Gauge
  Systems}.
\newblock 2009.

\bibitem{Itoyama2009}
H.~Itoyama, K.~Maruyoshi, and T.~Oota.
\newblock {The Quiver Matrix Model and 2d-4d Conformal Connection}.
\newblock {\em Prog. Theor. Phys.}, 123:957--987, 2010.

\bibitem{Bonelli2011}
G.~Bonelli, K.~Maruyoshi, A.~Tanzini, and F.~Yagi.
\newblock {Generalized matrix models and AGT correspondence at all genera}.
\newblock {\em JHEP}, 1107:055, 2011.

\bibitem{Maruyoshi2011}
K.~Maruyoshi and F.~Yagi.
\newblock {Seiberg-Witten curve via generalized matrix model}.
\newblock {\em JHEP}, 1101:042, 2011.

\bibitem{Schiappa2009}
R.~Schiappa and N.~Wyllard.
\newblock {An A\_r threesome: Matrix models, 2d CFTs and 4d N=2 gauge
  theories}.
\newblock 2009.

\bibitem{Mironov2009b}
A.~Mironov, A.~Morozov, and Sh. Shakirov.
\newblock {Matrix Model Conjecture for Exact BS Periods and Nekrasov
  Functions}.
\newblock {\em JHEP}, 02:030, 2010.

\bibitem{Mironov2010c}
A.~Mironov, A.~Morozov, and Sh. Shakirov.
\newblock {Conformal blocks as Dotsenko-Fateev Integral Discriminants}.
\newblock {\em Int. J. Mod. Phys.}, A25:3173--3207, 2010.

\bibitem{Mironov2010e}
A.~Mironov, Al. Morozov, and A.~Morozov.
\newblock {Conformal blocks and generalized Selberg integrals}.
\newblock {\em Nucl. Phys.}, B843:534--557, 2011.

\bibitem{Itoyama2010a}
H.~Itoyama and T.~Oota.
\newblock {Method of Generating q-Expansion Coefficients for Conformal Block
  and N=2 Nekrasov Function by beta-Deformed Matrix Model}.
\newblock {\em Nucl. Phys.}, B838:298--330, 2010.

\bibitem{Itoyama2011}
H.~Itoyama and N.~Yonezawa.
\newblock {$\epsilon$-Corrected Seiberg-Witten Prepotential Obtained From Half
  Genus Expansion in beta-Deformed Matrix Model}.
\newblock {\em Int. J. Mod. Phys.}, A26:3439--3467, 2011.

\bibitem{Mironov2010d}
A.~Mironov, A.~Morozov, and Sh. Shakirov.
\newblock {A direct proof of AGT conjecture at beta = 1}.
\newblock {\em JHEP}, 02:067, 2011.

\bibitem{Zhang2011}
Hong Zhang and Yutaka Matsuo.
\newblock {Selberg Integral and SU(N) AGT Conjecture}.
\newblock {\em JHEP}, 12:106, 2011.

\bibitem{Eguchi2010}
T.~Eguchi and K.~Maruyoshi.
\newblock {Seiberg-Witten theory, matrix model and AGT relation}.
\newblock {\em JHEP}, 07:081, 2010.

\bibitem{Eguchi2010a}
T.~Eguchi and K.~Maruyoshi.
\newblock {Penner Type Matrix Model and Seiberg-Witten Theory}.
\newblock (YITP-09-94), 2010.

\bibitem{Gaiotto2009a}
D.~Gaiotto.
\newblock {N=2 dualities}, April 2009.

\bibitem{Chekhov2002}
L.~Chekhov and A.~Mironov.
\newblock {Matrix models vs. Seiberg-Witten/Whitham theories}.
\newblock {\em Phys. Lett.}, B552:293--302, 2003.

\bibitem{Dyson1962}
F.~J. Dyson.
\newblock {Statistical theory of the energy levels of complex systems. I}.
\newblock {\em J. Math. Phys.}, 3:140--156, 1962.

\bibitem{Mironov2011a}
A.~Mironov, A.~Morozov, A.~Popolitov, and Sh. Shakirov.
\newblock {Resolvents and Seiberg-Witten representation for Gaussian
  beta-ensemble}.
\newblock 2011.

\bibitem{Nishinaka2011}
T.~Nishinaka and C.~Rim.
\newblock {$\beta$-deformed matrix model and Nekrasov partition function}.
\newblock 2011.

\bibitem{Maruyoshi2010}
K.~Maruyoshi and M.~Taki.
\newblock {Deformed Prepotential, Quantum Integrable System and Liouville Field
  Theory}.
\newblock {\em Nucl. Phys.}, B841:388--425, 2010.

\bibitem{Bonelli2011a}
G.~Bonelli, K.~Maruyoshi, and A.~Tanzini.
\newblock {Quantum Hitchin Systems via beta-deformed Matrix Models}.
\newblock 2011.

\bibitem{Gaiotto2009b}
D.~Gaiotto.
\newblock {Asymptotically free N=2 theories and irregular conformal blocks}.
\newblock August 2009.

\bibitem{Marshakov2009}
A.~Marshakov, A.~Mironov, and A.~Morozov.
\newblock {On non-conformal limit of the AGT relations}.
\newblock {\em Phys.Lett.B}, 682:125--129, 2009.

\bibitem{Itoyama2010}
H.~Itoyama, T.~Oota, and N.~Yonezawa.
\newblock {Massive Scaling Limit of beta-Deformed Matrix Model of Selberg
  Type}.
\newblock {\em Phys. Rev.}, D82:085031, 2010.

\bibitem{Chekhov2006}
L.~Chekhov and B.~Eynard.
\newblock {Matrix eigenvalue model: Feynman graph technique for all genera}.
\newblock {\em JHEP}, 12:026, 2006.

\bibitem{Chekhov2010a}
L.~Chekhov.
\newblock Logarithmic potential beta-ensembles and feynman graphs.
\newblock (ITEP/TH-33/10), September 2010.

\bibitem{Eynard2008a}
B.~Eynard and O.~Marchal.
\newblock {Topological expansion of the Bethe ansatz, and non- commutative
  algebraic geometry}.
\newblock {\em JHEP}, 03:094, 2009.

\bibitem{Chekhov2009}
L.~Chekhov, B.~Eynard, and O.~Marchal.
\newblock {Topological expansion of the Bethe ansatz, and quantum algebraic
  geometry}.
\newblock 2009.

\bibitem{Chekhov2010}
L.~O. Chekhov, B.~Eynard, and O.~Marchal.
\newblock {Topological expansion of beta-ensemble model and quantum algebraic
  geometry in the sectorwise approach}.
\newblock {\em Theor. Math. Phys.}, 166:141--185, 2011.

\bibitem{Klemm2008}
A.~Klemm and P.~Sulkowski.
\newblock {Seiberg-Witten theory and matrix models}.
\newblock {\em Nucl. Phys.}, B819:400--430, 2009.

\bibitem{Sulkowski2009}
P.~Sulkowski.
\newblock {Matrix models for $\beta$-ensembles from Nekrasov partition
  functions}.
\newblock {\em JHEP}, 04:063, 2010.

\bibitem{Poghossian2010}
R.~Poghossian.
\newblock {Deforming SW curve}.
\newblock {\em JHEP}, 04:033, 2011.

\bibitem{Fucito2011}
F.~Fucito, J.~F. Morales, D.~Ricci Pacifici, and R.~Poghossian.
\newblock {Gauge theories on Omega-backgrounds from non commutative
  Seiberg-Witten curves}.
\newblock {\em JHEP}, 05:098, 2011.

\bibitem{Huang2012}
Min xin Huang.
\newblock {On Gauge Theory and Topological String in Nekrasov-Shatashvili
  Limit}.
\newblock 2012.

\bibitem{Teschner2012}
J.~Teschner.
\newblock {Quantization of the Hitchin moduli spaces, Liouville theory, and the
  geometric Langlands correspondence I}, March 2012.

\bibitem{Zenkevich2011}
Y.~Zenkevich.
\newblock {Nekrasov prepotential with fundamental matter from the quantum spin
  chain}.
\newblock {\em Phys. Lett.}, B701:630--639, 2011.

\bibitem{Reffert2011}
S.~Reffert.
\newblock {General Omega Deformations from Closed String Backgrounds}.
\newblock {\em JHEP}, 04:059, 2012.

\bibitem{Hellerman2011}
S.~Hellerman, D.~Orlando, and S.~Reffert.
\newblock {String theory of the Omega deformation}.
\newblock {\em JHEP}, 01:148, 2012.

\bibitem{Hellerman2012}
S.~Hellerman, D.~Orlando, and S.~Reffert.
\newblock {The Omega Deformation From String and M-Theory}.
\newblock 2012.

\bibitem{Zamolodchikov1996}
A.~B. Zamolodchikov and Al.~B. Zamolodchikov.
\newblock {Structure Constants and Conformal Bootstrap in Liouville Field
  Theory}.
\newblock {\em Nuclear Physics B}, 477:577, 1996.

\bibitem{Goulian1990}
M.~Goulian and M.~Li.
\newblock {Correlation functions in Liouville theory}.
\newblock {\em Phys. Rev. Lett.}, 66:2051--2055, 1991.

\bibitem{Dotsenko1984}
V.~S. Dotsenko and V.~A. Fateev.
\newblock {Conformal algebra and multipoint correlation functions in 2D
  statistical models}.
\newblock {\em Nucl. Phys.}, B240:312, 1984.

\bibitem{Dotsenko1984a}
V.~S. Dotsenko and V.~A. Fateev.
\newblock {Four Point Correlation Functions and the Operator Algebra in the
  Two-Dimensional Conformal Invariant Theories with the Central Charge $c< 1$}.
\newblock {\em Nucl. Phys.}, B251:691, 1985.

\bibitem{Morozov2010}
A.~Morozov and Sh. Shakirov.
\newblock {The matrix model version of AGT conjecture and CIV-DV prepotential}.
\newblock {\em JHEP}, 08:066, 2010.

\bibitem{Nishinaka2012}
C.~Rim and T.~Nishinaka.
\newblock {Private communication}.

\bibitem{Morozov2012a}
A.~Morozov.
\newblock {Challenges of beta-deformation}.
\newblock 2012.

\bibitem{Mironov2012}
A.~Mironov, A.~Morozov, and Z.~Zakirova.
\newblock {Comment on integrability in Dijkgraaf-Vafa beta- ensembles}.
\newblock 2012.

\bibitem{Morozov2012}
A.~Morozov.
\newblock {Faces of matrix models}.
\newblock 2012.

\bibitem{Tai2010}
Ta-Sheng Tai.
\newblock {Uniformization, Calogero-Moser/Heun duality and Sutherland/bubbling
  pants}.
\newblock {\em JHEP}, 10:107, 2010.

\bibitem{Alday2009a}
L.~Alday, D.~Gaiotto, S.~Gukov, Y.~Tachikawa, and H.~Verlinde.
\newblock {Loop and surface operators in N=2 gauge theory and Liouville modular
  geometry}.
\newblock {\em JHEP}, 01:113, 2010.

\bibitem{Ferrari2012}
F.~Ferrari and M.~Piatek.
\newblock Liouville theory, n=2 gauge theories and accessory parameters,
  February 2012.

\bibitem{Wiegmann2005}
P.~Wiegmann and A.~Zabrodin.
\newblock {Large N expansion of the 2D Dyson gas}.
\newblock {\em J. Phys.}, A39:8933--8964, 2006.

\bibitem{Jurkiewicz1990}
J.~Jurkiewicz.
\newblock {REGULARIZATION OF THE ONE MATRIX MODELS}.
\newblock {\em Phys. Lett.}, B245:178--184, 1990.

\bibitem{Aganagic2011}
M.~Aganagic, M.~Cheng, R.~Dijkgraaf, D.~Krefl, and C.~Vafa.
\newblock {Quantum Geometry of Refined Topological Strings}.
\newblock 2011.

\bibitem{Galakhov2012}
D.~Galakhov, A.~Mironov, and A.~Morozov.
\newblock {S-duality as a beta-deformed Fourier transform}.
\newblock 2012.

\end{thebibliography}

\end{document}